\documentclass[sigconf]{acmart}

\usepackage{xcolor,colortbl}

\AtBeginDocument{%
  }

\copyrightyear{2025}
\acmYear{2025}
\setcopyright{cc}
\setcctype{by}
\acmConference[Websci '25]{17th ACM Web Science Conference}{May 20--24, 2025}{New Brunswick, NJ, USA}
\acmBooktitle{17th ACM Web Science Conference (Websci '25), May 20--24, 2025, New Brunswick, NJ, USA}
\acmDOI{10.1145/3717867.3717908}
\acmISBN{979-8-4007-1483-2/2025/05}

\usepackage{enumitem}

\newcommand{\rt}[1]{{\textcolor{black}{{#1}}}}

\usepackage{multirow}

\newcommand{\nsentences}{1296} 
\newcommand{\nreviews}{1043} 
\newcommand{\nmoviesword}{six} 
\newcommand{\nsentencesNoTBDs}{1286} 

\newcommand{\impact}{\textit{Impact}}
\newcommand{\sentiment}{\textit{Sentiment}}

\usepackage{newfloat}
\usepackage{listings}
\floatstyle{ruled}
\newfloat{listing}{tb}{lst}{}
\floatname{listing}{Listing}

\begin{document}

\title[Beyond the Lens]{Beyond the Lens: Quantifying the Impact of Scientific Documentaries through Amazon Reviews}



\author{Jill P. Naiman}
\affiliation{%
  \institution{University of Illinois}
  \city{Champaign}
  \country{USA}}
\email{jnaiman@illinois.edu}

\author{Aria Pessianzadeh}
\affiliation{%
  \institution{Drexel University}
  \city{Philadelphia}
  \country{USA}}

\author{Hanyu Zhao}
\affiliation{%
  \institution{Duke University}
  \city{Durham}
  \country{USA}}

\author{{AJ} Christensen}
\affiliation{%
  \institution{NASA Scientific Visualization Studio}
  \city{Baltimore}
  \country{USA}}

\author{Kalina Borkiewicz}
\affiliation{%
  \institution{The University of Utah}
  \city{Salt Lake City}
  \country{USA}}

\author{Shriya Srikanth}
\affiliation{%
  \institution{Harvard}
  \city{Cambridge}
  \country{USA}}

\author{Anushka Gami}
\affiliation{%
  \institution{University of Illinois}
  \city{Champaign}
  \country{USA}}

\author{Emma Maxwell}
\affiliation{%
  \institution{University of Illinois}
  \city{Champaign}
  \country{USA}}

\author{Louisa Zhang}
\affiliation{%
  \institution{University of Illinois}
  \city{Champaign}
  \country{USA}}

\author{Sri Nithya Yeragorla}
\affiliation{%
  \institution{University of Illinois}
  \city{Champaign}
  \country{USA}}

\author{Rezvaneh Rezapour}
\affiliation{%
  \institution{Drexel University}
  \city{Philadelphia}
  \country{USA}}
\email{sr3563@drexel.edu}







\renewcommand{\shortauthors}{Naiman et al.}

\newcommand{\avllong}{the Advanced Visualization Lab (AVL). This interdisciplinary team is housed at the National Center for Supercomputing Applications and consists of visualization designers who focus on cinematic presentations of scientific data}
\newcommand{\avlshort}{AVL}

\begin{abstract}
Engaging the public with science is critical for a well-informed population. A popular method of scientific communication is documentaries.
Once released, it can be difficult to assess the impact of such works on a large scale, due to the overhead required for in-depth audience feedback studies.
In what follows, we overview our complementary approach to qualitative studies through quantitative impact and sentiment analysis of Amazon reviews for several scientific documentaries.
In addition to developing a novel impact category taxonomy for this analysis, we release a dataset containing \nsentences\ human-annotated sentences from \nreviews\ Amazon reviews for \nmoviesword\ movies created in whole or part by \avllong. 
Using this data, we train and evaluate several machine learning and large language models, discussing their effectiveness and possible generalizability for documentaries beyond those focused on for this work. 
Themes are also extracted from our annotated dataset which, along with our large language model analysis, demonstrate a measure of the ability of scientific documentaries to engage with the public.
\end{abstract}




\maketitle

\section{Introduction}
Communicating science is an important goal of many societies and institutions \citep[e.g.,][]{national2017communicating}. 
Scientific documentaries have long served as an important communication medium for disseminating scientific knowledge to the general public. They bridge the gap between complex scientific concepts and public understanding, often influencing public opinion and policy on critical issues such as climate change, public health, and technology~\citep{barnes2008,punzo2015,vogt2016,borkiewicz2019cinematic,ytini}. 
Cinematic treatments of data visualization in the field of ``Cinematic Scientific Visualization'' (CSV) \citep{borkiewicz2022introducing} such as documentaries and IMAX films can be especially effective in engaging the general public \citep{dubeck2004fantastic,arroio2010,franconeri2021science,lee2022affective}.
From a media studies perspective, the narrative techniques and cinematographic strategies employed in documentaries can significantly enhance viewer engagement and retention of information~\cite{barrett2008assessing,yeo2018inconvenient,saputra2022impactful}. 
Research shows that the storytelling aspect of documentaries can lead to greater empathetic understanding and cognitive retention of scientific facts~\cite{ginting2024effects,gaunkar2022exploring}. The emotional responses elicited by documentaries are seen as key drivers for behavioral change and advocacy, influencing how individuals and communities respond to scientific challenges like environmental conservation or disease prevention~\cite{bieniek2019communicating}.

The impact of media was studied from the perspective of sociology, examining how documentaries reinforce or challenge cultural norms and societal structures~\cite{rezapour2017classification,diesner2015social}. The results show that documentaries can act as catalysts for social change by highlighting underrepresented issues and providing a platform for marginalized voices. This fosters a more inclusive public dialogue around scientific and technological advancements~\cite{bouzoubaa2024euphoria,conrad2022breaking,diesner2016assessing}.
These impacts extend beyond individual viewers to influence societal norms and values. By presenting scientific issues within relatable contexts, documentaries can alter public discourse, catalyze community actions, and even shift policy directions~\cite{atakav2024impact}. Moreover, the reach and accessibility of documentaries have been vastly expanded by advances in digital technology and online streaming platforms, which allow for unprecedented dissemination and engagement across varied global audiences.

Measuring the ``impact'' of scientific documentaries requires a multifaceted approach that integrates both qualitative and quantitative methods. 
Prior studies have demonstrated that filmmakers and designers of CSV presentations must navigate numerous creative and technical decisions to ensure their effectiveness \citep{woodward2015one,borkiewicz2022introducing,jensen2023evidence}.
Traditionally, assessing audience attitudes and cognition has relied on in-depth qualitative techniques such as interviews and thematic analysis, which, while insightful, can be labor-intensive and subject to bias \citep{chen2005top,cawthon2007effect,buck2013effect,fraser2012giant,smith2015aesthetics,smith2017capturing,jensen2023evidence,jenseninprep}. 
Additionally, the reliance on manual qualitative analysis in prior research has posed challenges related to efficiency and reproducibility \citep{bieniek2019communicating,atakav2024impact,barrett2008assessing}.

To address these limitations, quantitative approaches can serve as valuable complements, enhancing precision and scalability in impact assessments.
By integrating quantitative methods, researchers can overcome the constraints of qualitative-only studies, leading to a more comprehensive and rigorous evaluation of scientific media's influence.
With the emergence of digital platforms, viewers share their immediate reactions and detailed thoughts through online reviews, providing a rich dataset for analyzing public engagement and understanding~\cite [e.g.,][]{APPEL2016110}.
The interactivity provided by online platforms enhances the dissemination process, as viewers can discuss and share content, thereby amplifying the documentary's reach and impact. This engagement is measurable through the analysis of online reviews and social media commentary, which can serve as a feedback loop for content creators and scientists alike, indicating which aspects of their presentation meet the audience's needs and which do not~\cite{rezapour2017classification,bouzoubaa2024euphoria}.

We extend previous work by computationally analyzing the impact of scientific films using user-generated reviews and measuring the impact using a novel taxonomy of change and engagement in conjunction with a systematic approach to manual annotations. This approach employs natural language processing (NLP) techniques to systematically categorize and quantify viewer responses, thus providing a more scalable and objective measure of the documentaries' effectiveness in enhancing scientific literacy and influencing viewers' behavior. More specifically, we answer the following questions: 
 \setlist{nolistsep}
    \begin{itemize}[noitemsep]
    \item[-] {\textbf{RQ1:}} What are the most prevalent impacts of scientific documentaries on viewer cognition, attitudes, and interests? 
    \item[-] {\textbf{RQ2}:} What, if any, are some common themes in viewer cognition, attitude, and interests surrounding scientific documentaries?
\end{itemize}

To answer these questions, we extract reviews from Amazon for \nmoviesword\ prominent scientific documentaries and evaluate the tone, content, and thematic elements of viewer feedback. This data is further analyzed using our developed taxonomy, which categorizes impacts into ``Shift in Cognition (C)'', ``Attitudes Toward the Film (A)'', ``Interest with Science Topic (S)'', ``Impersonal report (I)'', and ``Not applicable (N)''.
Our data analysis shows that expressing different ``Attitudes Toward the Film'' is the most common \impact\ type reported in the reviews. 
The classification results suggest that large language model (LLM)-based classifiers, especially when the full context of the review is added to the prompt, are capable of successfully identifying \impact\ categories and their performance is generalizable to other datasets. Additionally, our thematic analysis of \impact\ groups demonstrates a diverse range of themes in the reviews. Viewers 
used mixed sentiments to show their attitude toward the film, while expressing environmental concerns or celebrating cosmic science were some of the themes related to shifts in cognition. 

Our paper makes the following contributions: (1) providing annotated data on impact, offering a dataset that has been carefully categorized and labeled to reflect the various types of influence that scientific documentaries have on public perception and interest\footnote{\url{https://huggingface.co/datasets/scidoc/websci2025}}; (2) introducing a novel taxonomy of engagement, which enables a structured analysis of how documentaries affect cognitive, affective, and interest dimensions of their audiences; and (3) employ advanced computational techniques to analyze viewer responses in a scalable and unbiased manner. 
These contributions enhance our understanding of the educational and societal impacts of scientific documentaries. The paper also provides a framework for future research to build upon, improving the effectiveness and reach of science communication through visual media.

\section{Related Work}
Social media platforms are invaluable resources for understanding public sentiment and behavior, offering real-time access to users' expressed opinions and emotions. By analyzing vast amounts of social media data, researchers can uncover trends, gauge audience reactions, and explore complex phenomena like consumer behavior and societal attitudes~\cite{elalaoui2018sentiment,cortis2021socialopinion}. Platforms like Twitter, Reddit, and YouTube provide large-scale datasets that enable studies to capture real-time feedback and public sentiment at an unprecedented scale in domains such as politics~\cite{ANTYPAS2023100242, Yarchi15032021,rezapour2017identifying}, healthcare~\cite{10.1007/978-3-031-27409-1_41,9810923}, and education~\cite{electronics11050715,luo2020like}.
For instance, sentiment analysis of tweets related to significant global events or media productions, such as the Netflix documentary Our Planet, reveals insights into public sentiment and media impact. This analysis helps creators and researchers understand audience engagement and the potential societal influence of content~\citep{ieeexplore_10533989}. Similarly, Reddit's discourse, known for its depth and variety, serves as a valuable source for exploring nuanced discussions and opinions~\citep{asurveyof2012opinion,10.1145/3543873.3587324,10.3389/frai.2023.1163577,bouzoubaa2024euphoria}.


Sentiment analysis has been central to leveraging social media data for understanding public opinion~\cite{ijerph15112537, QIAN2022103098}. Advances in NLP and machine learning have significantly improved sentiment classification tasks. Traditional approaches like Support Vector Machines (SVM) and K-Nearest Neighbors (KNN) have given way to deep learning methods, which excel in handling large, complex datasets~\cite{BANSAL2022100071}. Models such as Long Short-Term Memory (LSTM), Bidirectional LSTM (BiLSTM), and transformers like BERT and RoBERTa have achieved state-of-the-art performance in sentiment analysis~\citep{ieeexplore_9716923, ieeexplore_8629198, kapur2022sentimentanalysis}.
Recent comparisons of machine learning models for social media sentiment analysis reveal that deep learning approaches consistently outperform traditional algorithms. RoBERTa-LSTM hybrids, for instance, demonstrate exceptional accuracy and robustness in Twitter sentiment classification~\citep{ieeexplore_10331232}. These models' ability to understand the context and handle large datasets has made them indispensable for analyzing public opinion across diverse media platforms~\citep{arxiv_2408.08694, ieeexplore_10533989, arxiv_2407.13069}.

With the emergence of LLMs such as GPT and BERT-based architectures, sentiment analysis has evolved further, leveraging pre-trained models to understand complex linguistic patterns and contextual subtleties\cite{sayeed2023bert,chang2024survey}. These models enable transfer learning, allowing fine-tuning on specific domains to achieve superior performance compared to traditional and earlier deep learning methods~\cite{10.1145/3543873.3587324, arxiv_2407.13069}. 
Studies emphasize the importance of domain-specific tuning of NLP models. For example, the application of sentiment analysis to social media reviews of environmental documentaries or films not only provides feedback on content reception but also sheds light on broader societal attitudes towards environmental issues~\citep{acerbi2023sentiment}.
Opinion mining goes a step further than sentiment analysis by focusing on specific aspects of text that inform user perspectives. This fine-grained analysis has been applied to social media and online discourse, uncovering detailed insights into user opinions on products, services, or cultural phenomena~\citep{asurveyof2012opinion, kapur2022sentimentanalysis, Gerard_Botzer_Weninger_2023}.
For instance, \citep{kapur2022sentimentanalysis} employed BiLSTM combined with a random forest classifier to identify sentiments across multiple platforms, achieving high accuracy. 

Film reviews provide a specifically rich ground for sentiment analysis and opinion mining \cite{malini2019opinion}. They are rich in emotion, context, and critique, making them ideal for studying user opinions and their motivations. Existing research has predominantly focused on sentiment polarity, identifying whether reviews are positive, negative, or neutral. While effective, this binary/ternary classification often misses the granular details of user opinions~\citep{Mrabti2024AnEM, 10.5120/ijca2017916005}.
Researchers have called for extending these efforts to include opinion mining, which examines the thematic and contextual dimensions of reviews. For instance, \citep{rezapour2017classification} introduced ``Impact'' categories to analyze how specific aspects of films influence people's cognition and attitude. This nuanced approach revealed critical factors driving engagement, such as emotional resonance and storytelling quality. In addition, \cite{yue2019survey} emphasizes the importance of incorporating fine-grained sentiment analysis to move beyond simple polarity classification. This involves analyzing not only whether reviews are positive or negative but also understanding the rich contextual and emotional dimensions that shape user opinions. By capturing these subtleties, researchers can uncover deeper insights into audience preferences and motivations, ultimately enabling more tailored and effective content strategies.

Building on this body of work, 
by adopting a fine-grained approach, we seek to identify not only the overall sentiment but also the specific aspects of scientific films that shape public emotions and opinions. This holistic understanding will contribute to a deeper comprehension of user feedback, guiding content creators and industry stakeholders in optimizing their offerings.

\section{Data}

\subsection{Data Collection}

\newcommand{\avllonger}{a team of visualization designers who focus on cinematic presentations of scientific data (CSVT)}

A dataset of \nsentences\ sentences was collected from a total of \nreviews\ Amazon reviews of the \nmoviesword\ movies produced by \avllonger. \rt{The mean word count per review is 11.2 with a standard deviation of 9.0 words, and the mean word length across all reviews is 4.6 characters with a standard deviation of 2.7 characters.}
%
Movies include: SuperTornado: Anatomy of a MegaDisaster~\citep{superTornado2015}, Birth of Planet Earth~\citep{bope2019}, Solar Superstorms~\citep{solarsuperstorms2013,solarsuperstorms2015}, Seeing the Beginning of Time~\citep{sbot2017}, Space Junk~\citep{spacejunk2012}, and The Jupiter Enigma~\citep{jupterenigma2018}.   
In addition to the \nsentences\ collected for annotation and model training, 400 sentences were collected from Amazon reviews of the Hubble documentary~\citep{hubble2010} for final model evaluation.
Movies were selected in collaboration with the \avlshort\ as this team of visual designers expressed interest in quantifying the impact of their works.
Full reviews were downloaded using the \textsf{selectorlib} software package\footnote{\url{https://pypi.org/project/selectorlib/}}. 
In addition to the full text, the date and rating of the review were also collected. 

\subsection{Data Annotation}

\subsubsection{Taxonomy of Scientific Documentary Impact}
In seeking evidence for constructivist learning~\citep[e.g.,][]{bada2015constructivism}, we aim to track written language that indicates a statement about cognition or a statement about engagement \rt{(\autoref{tab:impact_categories})}.
As such, our impact categories are modeled off those used in the analysis of issue-focused documentaries \citep{rezapour2017classification} with several updates for our set of scientific documentaries
\citep[in Table 1,][]{rezapour2017classification}. 
More specifically, while internet comments are collected in too uncontrolled an environment to make any definitive conclusions about learning \citep[e.g.,][]{jensen_putting_2017}, we can identify trends among contributors in their comments associated with learning. Our impact category ``Shift in Cognition'' represents audiences' relative ability to identify terms, features, and concepts, up to and including metacognitive discussion of their own cognitive processes. 
    
In our impact categories scheme, we identify engagement through interest and attitudes (impact categories ``Attitudes Toward the Film'' and ``Interest with Science Topic''). Some measures of interest in films can be measured from collected comments. Although they are not conclusive proof of the existence of, or the strength of, a viewer's interest, patterns across large numbers of viewers can provide useful evidence. 
We use the category ``Interest with Science Topic'' when viewers make statements that evaluate the documentary content in its immediate context. 
The impact category ``Impersonal report'' is used when a viewer describes events in the film, and ``Not Applicable''; is the category used for sentences that do not have a specific relationship to the film or science topic.

Sentiment categories follow the typical three classes of ``Positive'', ``Neutral'' and ``Negative'' \citep[e.g.,][]{Liu2012}.
See \autoref{tab:impact_categories} for an overview of these definitions and examples of the combination of sentiment and impact categories used in this work.

\begin{table*}[t]
\resizebox{0.9\textwidth}{!}{
\begin{tabular}{ @{}p{1.4in}p{1.6in}p{1.6in}p{1.6in}p{1.6in}@{} }
\toprule 
\multicolumn{1}{c}{Impact Category} & \multicolumn{3}{c}{Sentiment Category}
\\\cmidrule(l){2-4}
 & \multicolumn{1}{c}{Positive (+)} & \multicolumn{1}{c}{Neutral (0)} & \multicolumn{1}{c}{Negative (-)} \\\midrule
Shift in Cognition (C)\rt{: ability to identify terms, features, and concepts, up to and including
metacognitive elements} & Person encountered a new idea or way of thinking & No change in way of thinking & Person expressed disagreement with scientific concepts presented in film \\ 
 & ``very informative about the development of tornados and the research behind their formation'' & ``We need to reign in the junk orbiting our planet before we lose something more important than the satellite Iridium 23, like a human life!'' & ``Unfortunately, HUGE lack of actual facts.'' \\
\hline
Attitudes Toward the Film (A)\rt{: evaluations of film technical and artistic attributes} & Person indicates positive attributes about film attributes & Person simply acknowledges film attributes & Person expresses dislike of film attributes \\ 
 & ``Way cool illustrations of our solar friend and how she can be a bad girl.'' & ``More technical than I thought it would be.'' & ``How could something so AWESOME and real life be made to be so boring?'' \\
\hline
Interest with Science Topic (S)\rt{: evaluations and connections to film scientific content} & Person expresses positive interest in science topic presented in film & Person expresses a connection with the science topic & Person expresses disinterest in science topic \\ 
 & ``Space intrigues me greatly, and this was an amazing program.'' & ``I've followed supercomputing since inception (and currently run some of their software).'' &  ``used to intrigue me as a child, but knowing reality now, I seek real answers over lies.''
 \\
\hline
Impersonal report (I)\rt{: descriptions of events in a film without evaluation} & Person describes events in the film with positive connotations & Person describes events in the film & Person describes events in the film with negative connotations \\ 
 & ``This video reveals the cutting edge science gathered by the Juno Mission to Jupiter.'' & ``There's not much footage of the Joplin Tornado itself, most of this is footage of the wreckage.'' & ``When there were other scenes to show a specific class of Space Junk then what you saw was simply a different geometric shape and that is it.'' \\
\hline 
Not Applicable (N)\rt{: statements with no specific relation to film or science topic} & Positive comment, but no specific relation to film or topic & Comment has no specific relation to film or topic & Negative comment, but no specific relation to film or topic  \\ 
 & ``Interesting'' & ``This has never happened before.'' & ``Documentaries have tried to be too interesting lately.'' \\
\bottomrule
\end{tabular}}
\caption{Sentiment and impact categories with definitions and examples.}
\label{tab:impact_categories}
\end{table*}

\subsection{Annotation Procedure}
After reviewing a sample of reviews, we noticed that individual sentences within a review can address different impacts and may express opposite sentiments. As a result, we opted to perform annotations at the sentence level. We used the Zooniverse\footnote{\url{https://www.zooniverse.org/}} citizen science platform, one of the largest platforms for non-professionals to participate in the scientific process, for the annotation process. 

Using Zooniverse's default interface for textual data, we displayed a single sentence from a review followed by the sentence within the full review for context.
This display and the first stage of the annotation process is depicted in \autoref{fig:first_step_annotation} in the Appendix in which the user is asked to select their first choice for the sentiment category (Positive, Negative, or Neutral).

Annotation of each of the \nsentences\ sentences occured in batches of $\approx$200 sentences, with each sentence in a batch ``retired'' once three users have completed annotations on the sentence.
The breakdown of the number of annotated sentences per movie is shown in \autoref{tab:movie_info}.


\subsection{Data Cleaning and Analysis} \label{sec:datacleaning}
After completing the annotation on \sentiment\ and \impact, we used Cohen's Kappa \citep{cohen1960coefficient} to measure the degree of reliability and agreement between the annotators. For sentiment categories, the kappa score was relatively consistent across each category with 0.66 for ``Positive'' sentiments, 0.67 for ``Neutral'' sentiments, and 0.68 for ``Negative'' sentiments.
For impact categories, the lowest Kappa measurement was for ``Impersonal Report'' (0.61) with higher measurements for ``Interest in Science Topic'', ``Shift in Cognition'' and ``Not applicable'' (0.67) along with ``Attitudes Toward the Film'' (0.68).

\begin{figure}[t]
\centering
\includegraphics[width=0.7\columnwidth]{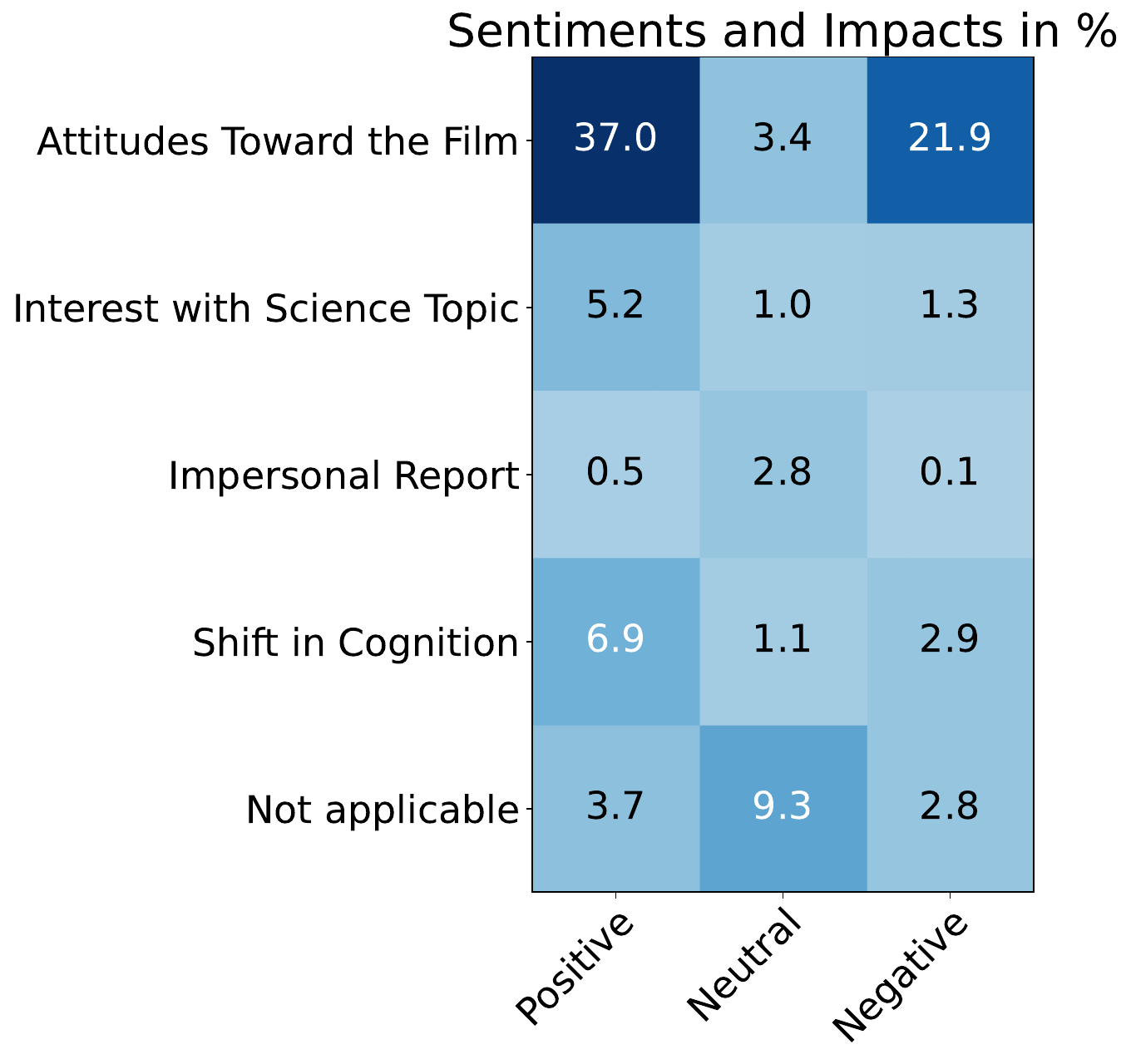} 
\caption{Breakdown of full \nsentencesNoTBDs\ sentences in the annotated dataset as described in the ``Data'' section.  Numbers are percentages of the total sentences in the dataset within each Sentiment and Impact combination.}
\label{fig:sentiment_impact}
\end{figure}

Final sentiment and impact categories were decided by a majority vote. In $\approx$10\% of the sentences, where no majority consensus was possible, all annotators discussed each sentence in a group and either a majority consensus was reached, or the sentence was marked as ``no agreement''. This resulted in eight ``no agreement'' sentences, which we excluded from our analysis.
Finally, we excluded sentences that were only emojis or consisted of solely punctuation marks (e.g., ``:D'').
The composition of the remaining \nsentencesNoTBDs\ sentences in sentiment and impact categories is shown as a percentages heatmap in \autoref{fig:sentiment_impact} with the per-media breakdown in raw numbers of sentiments of sentiment and impact category per film shown in \autoref{tab:movie_info}.


\begin{table*}[h]
\begin{center}
\begin{tabular} {@{}lcccccccccc@{}} 
\toprule
\multicolumn{1}{c}{Media} & \multicolumn{1}{c}{Year} & \multicolumn{1}{c}{\# Sentences} & \multicolumn{3}{c}{Sentiment [\%]} & \multicolumn{5}{c}{Impact [\%]} \\\cmidrule(lr){4-6} \cmidrule(l){7-11}
\multicolumn{1}{c}{} & \multicolumn{1}{c}{} & \multicolumn{1}{c}{} & \multicolumn{1}{c}{+} & \multicolumn{1}{c}{0} & \multicolumn{1}{c}{-} & \multicolumn{1}{c}{A} & \multicolumn{1}{c}{S} & \multicolumn{1}{c}{C} & \multicolumn{1}{c}{I} &  \multicolumn{1}{c}{N} \\\cmidrule(lr){1-3}\cmidrule(lr){4-6} \cmidrule(l){7-11}

Space Junk & 2012 & 137 & \cellcolor[rgb]{0.37673202614379087,0.6530718954248366,0.8224836601307189}53.28 & \cellcolor[rgb]{0.8672664359861592,0.9193540945790081,0.967520184544406}13.14 & \cellcolor[rgb]{0.6718954248366014,0.8143790849673203,0.9006535947712418}33.58 & \cellcolor[rgb]{0.8605767012687427,0.29554786620530565,0.011118800461361017}73.72 & \cellcolor[rgb]{0.9991387927720108,0.9478662053056517,0.8965936178392926}2.92 & \cellcolor[rgb]{0.9982775855440216,0.9349480968858132,0.8716186082276047}5.84 & \cellcolor[rgb]{0.9982775855440216,0.9349480968858132,0.8716186082276047}5.84 & \cellcolor[rgb]{0.9964321414840446,0.9072664359861592,0.8181007304882737}11.68 \\ 
Solar Superstorms$^\dagger$ & 2013(15) & 107 & \cellcolor[rgb]{0.28089196462898885,0.5876201460976547,0.7850826605151865}60.75 & \cellcolor[rgb]{0.8318339100346022,0.8957324106113033,0.9557093425605536}17.76 & \cellcolor[rgb]{0.8023068050749712,0.8760476739715494,0.9458669742406767}21.5 & \cellcolor[rgb]{0.9515570934256055,0.4311418685121107,0.09657823913879277}60.75 & \cellcolor[rgb]{0.9988927335640139,0.9441753171856978,0.8894579008073817}3.74 & \cellcolor[rgb]{0.996555171088043,0.9091118800461361,0.8216685890042291}11.21 & \cellcolor[rgb]{0.9982775855440216,0.9349480968858132,0.8716186082276047}5.61 & \cellcolor[rgb]{0.9942176086120723,0.8610226835832372,0.7259669357939253}18.69 \\ 
SuperTornado & 2015 & 555 & \cellcolor[rgb]{0.36159938485198,0.6427374086889658,0.8165782391387928}54.41 & \cellcolor[rgb]{0.8377393310265283,0.8996693579392541,0.9576778162245291}17.12 & \cellcolor[rgb]{0.7358708189158016,0.8415686274509804,0.923044982698962}28.47 & \cellcolor[rgb]{0.9515570934256055,0.4311418685121107,0.09657823913879277}60.72 & \cellcolor[rgb]{0.9979084967320262,0.9294117647058824,0.8609150326797386}7.03 & \cellcolor[rgb]{0.9961860822760477,0.9035755478662053,0.810965013456363}12.43 & \cellcolor[rgb]{0.9990157631680123,0.9460207612456748,0.8930257593233372}3.42 & \cellcolor[rgb]{0.9949557862360631,0.8772625913110342,0.7584467512495194}16.4 \\ 
Seeing the Beginning of Time & 2017 & 233 & \cellcolor[rgb]{0.35151095732410614,0.6358477508650519,0.812641291810842}55.36 & \cellcolor[rgb]{0.8554555940023069,0.9114801999231065,0.9635832372164552}14.59 & \cellcolor[rgb]{0.7161860822760477,0.8332026143790849,0.916155324875048}30.04 & \cellcolor[rgb]{0.898961937716263,0.3483275663206459,0.039907727797001157}68.67 & \cellcolor[rgb]{0.9966782006920415,0.9109573241061131,0.8252364475201845}10.73 & \cellcolor[rgb]{0.9976624375240293,0.9257208765859285,0.8537793156478277}7.73 & \cellcolor[rgb]{1.0,0.9607843137254902,0.9215686274509803}0.0 & \cellcolor[rgb]{0.9960630526720492,0.9016224529027297,0.8071664744329105}12.88 \\ 
The Jupiter Enigma & 2018 & 219 & \cellcolor[rgb]{0.491764705882353,0.7219684736639754,0.8547789311803152}45.66 & \cellcolor[rgb]{0.7703191080353711,0.8562091503267973,0.9351018838908113}25.57 & \cellcolor[rgb]{0.7309496347558632,0.8394771241830065,0.9213225682429834}28.77 & \cellcolor[rgb]{0.9855132641291812,0.5330103806228376,0.2125951557093427}51.6 & \cellcolor[rgb]{0.996555171088043,0.9091118800461361,0.8216685890042291}10.96 & \cellcolor[rgb]{0.9960630526720492,0.9016224529027297,0.8071664744329105}12.79 & \cellcolor[rgb]{0.9985236447520185,0.938638985005767,0.8787543252595156}5.02 & \cellcolor[rgb]{0.9938485198000769,0.8529027297193387,0.7097270280661284}19.63 \\ 
Birth of Planet Earth & 2019 & 35 & \cellcolor[rgb]{0.4019530949634756,0.6702960399846213,0.8323260284505959}51.43 & \cellcolor[rgb]{0.8584083044982699,0.9134486735870818,0.9645674740484429}14.29 & \cellcolor[rgb]{0.6620530565167244,0.8101960784313725,0.8972087658592849}34.29 & \cellcolor[rgb]{0.8782929642445213,0.31990772779700116,0.024405997693194924}71.43 & \cellcolor[rgb]{0.9991387927720108,0.9478662053056517,0.8965936178392926}2.86 & \cellcolor[rgb]{0.9955709342560554,0.8907958477508651,0.7855132641291811}14.29 & \cellcolor[rgb]{1.0,0.9607843137254902,0.9215686274509803}0.0 & \cellcolor[rgb]{0.9964321414840446,0.9072664359861592,0.8181007304882737}11.43 \\

\hline
\end{tabular}
\end{center}
\footnotesize{$^\dagger$ Solar Superstorms sentences come from two media presentations: Solar Superstorms - a TV Episode from ``Cosmic Journeys'' \citep{solarsuperstorms2013} and a full-length documentary Solar Superstorms: Journey to the Center of the Sun \citep{solarsuperstorms2015}.}\\
\caption{Total number of annotated sentences per media.  Single character sentiment and impact codes are defined in \autoref{tab:impact_categories}.}
\label{tab:movie_info}
\end{table*}

\section{Methodology}

\subsection{Classification Models} \label{section:classification_models}
We leveraged three types of classifiers for sentiment and impact classification: (1) three traditional machine learning algorithms, (2) two transformer-based models, and (3) five large language models. 
To create a train and test set, we split the data into 70\% training and 30\% test sets. For transformer-based models, we use 20\% of the training data for validation.
Stratified data split was used based on \impact\ labels to split the data.
We measured the classification performance on the test set after training each model using precision, recall, and F1 scores.

\paragraph{Baseline model: }
We applied three baseline models, i.e., Support Vector Machine (SVM), Logistic Regression, and Decision Tree Classifier. 
We used Term Frequency-Inverse Document Frequency (TF-IDF) to vectorize the sentences. 

\paragraph{Transformer-based models: }
We leveraged two transformer-based models, i.e., BERT \cite{DBLP:journals/corr/abs-1810-04805} and RoBERTa \cite{DBLP:journals/corr/abs-1907-11692}, to classify sentences. Models were fine-tuned using pre-trained weights from bert-base-uncased and roberta-base, respectively. The sentences were tokenized using the corresponding tokenizers (BertTokenizer or RobertaTokenizer).
Fine-tuning was performed using the Hugging Face API\footnote{https://huggingface.co/}, and hyperparameters were set to three epochs, a batch size of 8, a learning rate scheduler with 500 warmup steps, and a weight decay of 0.01. Model evaluation occurred at the end of each epoch using the validation set.

\paragraph{LLMs: }
We used three closed-weight models from OpenAI: GPT3.5 (gpt-3.5-turbo-0125) \cite{openai2023chatgpt}, GPT4 (gpt-4-turbo-2024-04-09) \cite{openai2023gpt4}, GPT4o (gpt-4o-2024-08-06), and two open-weight LLMs: llama3*8b (llama3-8b-8192) \cite{llama3modelcard} and Mixtral 8*7B (mixtral-8x7b-32768) \cite{jiang2024mixtral} for classification in two different ways: with and without the context (full review) as the textual context for the target sentence.
In the first method (``w/o context''), the model is instructed to predict the correct label for the target sentence based on the definitions provided for each class. 
This is akin to having the model classify on only the top sentence (without the ``Full review'') depicted in \autoref{fig:first_step_annotation}.
In the second approach (``with context''), the full review is included in the prompt to guide the model in predicting the correct label for the target sentence. This is similar to providing the ``Full review'' shown in \autoref{fig:first_step_annotation} and indicating the target sentence.
For the LLM-based classifiers, we used the test set to evaluate the models' performance as well as the efficiency of each prompting method.
Temperature was set to 0 for all model experiments for maximum consistency.

\paragraph{Fine-tuned LLMs: }
To explore the helpfulness of additional fine-tuning for the classification purpose, we finetuned GPT4o (gpt-4o-2024-08-06) using the training set.
After fine-tuning, we repeated the classification task with the same prompts to predict the label for each sentence. The performance was then evaluated on the test set to assess the improvement compared to the base model.

\subsection{Thematic Analysis}
To develop a richer understanding of the contextual nuances related to each impact, we employed thematic analysis to extract themes and patterns associated with each impact category. For this purpose, we conducted an in-depth analysis of sentences tagged with each impact category using LLooM \cite{lam2024concept}, an advanced framework for concept induction powered by LLMs.

\section{Results}

\subsection{Classification} \label{section:classification}

Table \ref{tab:Classification Result Sentiment} shows the performance of different classifiers on our dataset. Transformer- and LLM-based classifiers outperformed baseline models. GPT4 and GPT4o with context embedded in the prompts are performing the best compared to similar models, with F1 scores of 0.71 (GPT4o) for \textit{Impact} and 0.85 (GPT4) for \textit{Sentiment}, respectively. Our fine-tuned model of GPT4o resulted in the highest performance for both tasks, 0.76 and 0.88. This confirms the usefulness of fine-tuning LLMs with additional data. We also observed that while using full reviews as context is improving the performance of Sentiment classifiers, its helpfulness for Impact classification seems less significant. 

\begin{table}[b]
\centering
\resizebox{0.9\columnwidth}{!}{%
\begin{tabular}{@{}cllll|lll@{}}
\toprule
\multicolumn{1}{l}{} &  & \multicolumn{3}{c}{Impact} & \multicolumn{3}{c}{Sentiment} \\ \cmidrule(lr){3-5} \cmidrule(l){6-8}
\multicolumn{1}{l}{} &  & P & R & F1 & P & R & F1 \\\cmidrule(lr){1-2}\cmidrule(lr){3-5} \cmidrule(l){6-8}
\multirow{3}{*}{} & SVM & 0.57 & 0.61 & 0.58 & 0.68 & 0.65 & 0.6 \\
 & Logistic Regression & 0.56 & 0.62 & 0.57 & 0.63 & 0.64 & 0.63 \\
 & Decision Tree & 0.47 & 0.61 & 0.51 & 0.49 & 0.55 & 0.49 \\\midrule
\multirow{2}{*}{} & BERT & 0.56 & 0.67 & 0.59 & 0.75 & 0.77 & 0.75 \\
 & RoBERTa & 0.62 & 0.72 & 0.66 & 0.76 & 0.76 & 0.76 \\\midrule
\multirow{5}{*}{\rotatebox{90}{w/o context}} & GPT3.5 & 0.73 & 0.65 & 0.67 & 0.76 & 0.77 & 0.76 \\
 & GPT4 & 0.75 & 0.68 & 0.69 & 0.82 & 73 & 0.75 \\
 & GPT4o & 0.76 & 0.68 & 0.7 & 0.82 & 0.7 & 0.72 \\
 & LLama & 0.68 & 0.45 & 0.47 & 0.79 & 0.71 & 0.73 \\
 & Mixtral & 0.73 & 0.52 & 0.57 & 0.81 & 0.74 & 0.76 \\\midrule
\multirow{5}{*}{\rotatebox{90}{with context}} & GPT3.5 & 0.68 & 0.64 & 0.61 & 0.77 & 0.8 & 0.77 \\
 & GPT4 & 0.73 & 0.68 & 0.69 & 0.86 & 0.84 & 0.85 \\
 & GPT4o & 0.74 & 0.71 & 0.71 & 0.84 & 0.78 & 0.8 \\
 & LLama & 0.67 & 0.57 & 0.6 & 0.8 & 0.8 & 0.8 \\
 & Mixtral & 0.73 & 0.63 & 0.62 & 0.82 & 0.8 & 0.81 \\\midrule
\multicolumn{1}{l}{} & GPT4o Fine-tuned & \textbf{0.77} & \textbf{0.77} & \textbf{0.76} & \textbf{0.89} & \textbf{0.87} & \textbf{0.88} \\ \bottomrule
\end{tabular}%
}
\caption{Classification results for sentiment and impact categories (P: Precision, R: Recall, F1: F1-Score)}
\label{tab:Classification Result Sentiment}
\end{table}

\subsection{Error Analysis}
To enhance our understanding of our top-performing models, we conducted a detailed analysis of the fine-tuned GPT4o model based on the \impact\ class, which is the central focus of our study. \autoref{tab:F1 per class} shows the F1 score per impact class \rt{(center column)}. ``Attitudes Toward the Film'' has the highest classification performance of 0.86 while the F1 score for ``Impersonal Report'' is lowest at 0.42. 
%
\rt{These trends rightly align with the Cohen's Kappa measure of human annotator intercoder agreement \citep{mchugh2012interrater} shown in the right column of \autoref{tab:F1 per class}.  While most categories show substantial agreement between annotators ($\kappa > 0.6$), ``Impersonal Report'' is on the borderline between moderate and substantial agreement.}
\rt{The relatively larger} disparity \rt{in F1 score from the model's classifications when compared to differences in Cohen's Kappa} may be linked to the imbalance of impact categories in the dataset (as seen across the entire dataset in \autoref{fig:sentiment_impact} and \autoref{tab:movie_info}) during fine-tuning process or the more challenging nature of identifying this category. 

\autoref{tab:confusion matrix} demonstrates the confusion matrix of \impact\ classes, where we can see the most common type of misclassifications by the model. It is evident that ``Attitudes Toward the Film'' and ``Impersonal Report'' are most often misclassified as ``Shift in Cognition'', while ``Interest with Science Topic'', ``Not applicable'', and ``Shift in Cognition'' are mostly misclassified as ``Attitudes Toward the Film''. 
\begin{table}[b]
\centering
\resizebox{\columnwidth}{!}{%
\begin{tabular}{@{}lccccc@{}}
\hline
\textbf{True Label} & \multicolumn{5}{c@{}}{\textbf{Predicted Label}}\\
\cmidrule(l){2-6}
& Attitudes &  Impersonal & Interest & NA & Shift  \\ 
\midrule
  Attitudes Toward the Film  & 202 & 1 & 4 & 5 & 17  \\
  Impersonal Report  & 3 & 4 & 0 & 1 & 4  \\
  Interest with Science Topic  & 10 & 0 & 11 & 1 & 4  \\
  Not applicable & 20 & 2 & 0 & 31 & 4  \\
  Shift in Cognition  & 8 & 0 & 1 & 1 & 32  \\
\hline
\end{tabular}}

\caption{Confusion Matrix for True and Predicted Impact Categories: Attitudes (Attitudes Toward the Film), Impersonal (Impersonal Report), Interest (Interest with Science Topic), NA (Not applicable), and Shift (Shift in Cognition)}
\label{tab:confusion matrix}

\end{table}

\begin{table}[t]

\begin{tabular}{@{}lcc@{}}
\toprule
\rt{Impact Category} & \rt{F1} & \rt{Cohen's Kappa} \\
\hline
Attitudes Toward the Film & 0.86 & \rt{0.68} \\ 
Impersonal Report & 0.42 & \rt{0.61} \\ 
Interest with Science Topic & 0.52 & \rt{0.67} \\ 
Shift in Cognition & 0.62 & \rt{0.67} \\ 
Not applicable & 0.65 & \rt{0.67} \\ \bottomrule
\end{tabular}
\caption{F1 score \rt{(model) and Cohen's Kappa (human-annotations)} per class for Impact.}
\label{tab:F1 per class}
\end{table}

\subsection{Generalizability Test} \label{section:generalizability}
To evaluate the generalizability of our models in successfully capturing the impact of movies, we applied the best model, with the highest classification performance (``GPT4o Fine-tuned'') to the Hubble dataset and predicted the labels for both \impact\ and \sentiment. 
\autoref{tab:Hubble_Prediction} shows the distribution of predicted classes for this dataset. While the majority of reviews show ``Positive'' sentiment toward the Hubble documentary, ``Attitudes Toward the Film'' is the most frequent type of impact predicted by our model in the movie review. 
This aligns with the prevalence of the ``Attitudes Toward the Film'' impact category among the other movies in 
 \autoref{tab:movie_info}.

To assess the model's performance, we tasked two annotators with evaluating the accuracy of the predicted labels for both the \impact\ and \sentiment\ categories. The mean agreement for \impact\ is approximately 85.3\%. For \sentiment\ agreement is $\sim$87.6\%, demonstrating a substantial level of alignment between the annotators and the classification model.

\begin{table}[b]
\centering
\begin{tabular}{@{}lc|lc@{}}
\toprule
\textbf{Impact} &  & \textbf{Sentiment} & \\ \cmidrule(l){1-2}\cmidrule(l){3-4}
Attitudes Toward the Film & 307 & Positive & 250 \\
Not applicable & 50 & Neutral & 94 \\
Interest with Science Topic & 17 & Negative & 56 \\
Shift in Cognition & 15 &  &  \\
Impersonal Report & 11 &  & \\\bottomrule
\end{tabular}
\caption{Distribution of predicted classes for Hubble dataset}
\label{tab:Hubble_Prediction}

\end{table}

\subsection{Thematic Analysis}
\autoref{tab:Themes Impact} shows the themes of discussions in sentences, labeled with each impact category using LLooM. We exclude ``Not applicable'' from \autoref{tab:Themes Impact} as this category covers a wide variety of comments which are not relevant to the overall analysis of the movies' impacts.
Each \impact\ contains several related themes based on how reviewers have described the movie. While sentences labeled as ``Attitudes Toward the Film'' discuss various positive, negative, or mixed feedback on the movie, e.g., \textit{``Kind of boring bearly any action'' or ``The graphics are... OUT OF THIS WORLD''}, reviews tagged with ``Shift in Cognition'' cover themes such as space science and environmental concerns or information quality, e.g., \textit{``Jupiter is a stunningly gorgeous planet, and it's amazing to finally learn some of its secrets.''}

Context around ``Interest with Science Topic'' includes mentions of science and technical aspects, emotional and experiential reactions, content quality, and learning, e.g., \textit{``If observing and learning of the universe is your thing, you will love this''}. ``Impersonal Report'' often consists of safety issues or historical and informational aspects of the movies, e.g., \textit{``An 1859 type event will occur again and cause
1 to 2 Trillion dollars in damage.''} These themes collectively show how the viewers have interacted with scientific documentaries and how these movies engage with their cognition, attitudes, and interests.

\begin{table*}[t]
\centering
\resizebox{0.9\textwidth}{!}{%
\begin{tabular}{@{}lll@{}}
\toprule
Impact class & Themes & Examples \\
\midrule
\multirow{4}{*}{Attitudes Toward the Film} & \begin{tabular}[c]{@{}l@{}}Positive Engagement and Impact e.g. Positive feedback, \\ Desire for more similar content, \\ Describing the movie as engaging \& informative\end{tabular} & Very interesting and informative, liked it a lot. \\ \cline{2-3}
& \begin{tabular}[c]{@{}l@{}}Negative Feedback and Criticism e.g. Poor narrative, \\ lack of excitement, poor production or unengaging experience\end{tabular} & Kind of boring bearly any action \\ \cline{2-3}
& \begin{tabular}[c]{@{}l@{}}Content Features e.g. Visuals and graphics, Technical \\ and scientific features, Historical or factual aspects of the movie\end{tabular} & The graphics are... OUT OF THIS WORLD. \\ \cline{2-3}
& Mixed Opinions or Doubt e.g. Skepticism or criticism & \begin{tabular}[c]{@{}l@{}}The quality of this film is very good, but someone \\ should have checked for misinformation in the script.\end{tabular} \\ \midrule

\multirow{3}{*}{Shift in Cognition} & \begin{tabular}[c]{@{}l@{}}Space and Cosmic Science e.g. Orbital Mechanics, \\ Cosmic Learning\end{tabular} & \begin{tabular}[c]{@{}l@{}}Jupiter is a stunningly gorgeous planet, and \\ it's amazing to finally learn some of its secrets.\end{tabular} \\ \cline{2-3}
& \begin{tabular}[c]{@{}l@{}}Weather and Environmental Concerns e.g. Structural Safety, \\ Weather Knowledge, Environmental Impact\end{tabular} & \begin{tabular}[c]{@{}l@{}}It's about a terrible disaster to a community, \\ but also how they came back together.\end{tabular} \\ \cline{2-3}
& \begin{tabular}[c]{@{}l@{}}Criticism and Information Quality e.g. Misleading or Incorrect,\\  Lack of Clarity or New Insights\end{tabular} & \begin{tabular}[c]{@{}l@{}}I gave this 2 stars because they had \\ some facts wrong.\end{tabular} \\ \midrule

\multirow{4}{*}{Interest with Science Topic} & \begin{tabular}[c]{@{}l@{}}Science and Technical Aspects e.g. Technical Expertise, \\ Weather Science, Science Exploration, Space Interest\end{tabular} & \begin{tabular}[c]{@{}l@{}}If observing and learning of the universe \\ is your thing, you will love this.\end{tabular} \\ \cline{2-3}
& \begin{tabular}[c]{@{}l@{}}Emotional and Experiential Reactions e.g. Uncertainty or \\ Suspense, Emotional Experience, Relaxation or Fatigue\end{tabular} & \begin{tabular}[c]{@{}l@{}}The first documentary leaves us hanging, will \\ Juno be able to do what its designers want it to do\end{tabular} \\ \cline{2-3}
& \begin{tabular}[c]{@{}l@{}}Content Quality and Presentation e.g. Media Coverage, \\ Storytelling, Interesting Content, Critical Perspective\end{tabular} & Best of all, it is based upon a true story. \\ \cline{2-3}
& \begin{tabular}[c]{@{}l@{}}Learning and Engagement e.g. Informative Enjoyment, \\ Progress and Improvement\end{tabular} & \begin{tabular}[c]{@{}l@{}}It's an interesting movie subject wise but if it \\ had better 3D, I be drawn to see it again.\end{tabular} \\ \midrule

\multirow{3}{*}{Impersonal Report} & Tornado and Safety e.g. Building Safety, Tornado Consequences & \begin{tabular}[c]{@{}l@{}}This documentary describes positive building \\ code changes to help avert total demolition in a \\ tornado like this.\end{tabular} \\ \cline{2-3}
& \begin{tabular}[c]{@{}l@{}}Space and Astronomy e.g. Space Debris, \\ Solar Phenomena, Planetary Systems\end{tabular} & \begin{tabular}[c]{@{}l@{}}An hour of video images of solar observatory \\ and computer simulations of solar process and flares.\end{tabular} \\ \cline{2-3}
& \begin{tabular}[c]{@{}l@{}}Historical and Informational Content e.g. \\ Historical Events, Interviews and Reports\end{tabular} & \begin{tabular}[c]{@{}l@{}}An 1859 type event will occur again and cause \\ 1 to 2 Trillion dollars in damage\end{tabular} \\ \bottomrule
 
\end{tabular}}
\caption{Themes related to each Impact category}
\label{tab:Themes Impact}

\end{table*}

\section{Discussion}

In conversation with the \avlshort\, several of the interests the group emphasized were (1) a desire to see a breakdown between positive and negative views of their documentaries, (2) an understanding of what film attributes might drive a viewer's opinion of the film, and (3) that, as often engagement in and of itself is a goal, what measurements of audience engagement might be evident in comments.
These interests align with our two research questions.

\vspace{0.2cm}
\noindent\textbf{RQ1: The most prevalent impacts of scientific documentaries. } 
Our findings highlight the substantial role that scientific documentaries play in shaping audience engagement and attitudes. The predominance of positive sentiments, particularly those linked to ``Attitudes Toward the Film'' (37\% of all sentences), suggests that these films successfully resonate with their audiences. This is likely due to the compelling storytelling and production quality typical of impactful documentaries. This supports existing literature on the ability of documentary films to evoke emotional responses and create a sense of connection between viewers and experts \cite{nisbet2009documentary,corner2002performing}.

The disparity between ``Positive''/``Attitudes Toward the Film'' (37\%) and ``Negative''/``Attitudes Toward the Film'' (21.9\%) further underscores a general trend of audience favorability. This aligns with findings that highlight how carefully crafted documentaries can frame scientific content in ways that promote optimism, curiosity, and a sense of wonder about the natural or scientific world \cite{gregory1998science}. The approximately $\sim$1.7 greater prevalence of positive over negative sentiments reflects the potential of documentaries to act as persuasive and motivational tools, aligning with interests (1) and (3) of the \avlshort, focusing on emotional resonance and knowledge dissemination.
``Interest with Science Topic'' (5.2\%) and ``Shift in Cognition'' (6.9\%) prevalence, as significant secondary impacts, indicate that these films go beyond entertainment, fostering intellectual engagement and engaging with viewer perspectives. This mirrors findings from studies on science communication that emphasize the dual purpose of documentaries: sparking curiosity and promoting critical thinking \cite{dahlstrom2014using}.
The fact that these impact types, although not the most prevalent, still contribute to a notable portion of the dataset shows that documentaries can serve as informal educational tools, stimulating deeper engagement with scientific topics and potentially influencing public understanding of science.

\vspace{0.2cm}
\noindent\textbf{RQ2: Common themes and discourse in viewers' feedback.}
As indicated in \autoref{tab:Themes Impact}, there are several themes that align with each impact category.  For example, attitudes toward the film which lean negative often mention poor narration or production quality or a lack of excitement, while positive attitudes mention a desire for similar content and that the content is engaging and informative.  These thematic trends align with interest (2) of the \avlshort. 

Additionally, many themes listed in \autoref{tab:Themes Impact} involve engagement as well as the visualization/technical aspects of the films, aligning with the interest (3) of the \avlshort. 
Not only is engagement talk more evidence of learning, but it is also a building block that can lead to behaviors with prosocial impact. These behaviors include de-stigmatizing and sharing science information with the community in informal environments, improving personal science identity (which can lead to the growth of future science professionals), and advocating for more research funding, political support, and science-motivated decision-making \citep[e.g.,][]{lee_robbins_affective_2022}.

The themes identified further highlight the dual role of media and documentary films: as both educational tools and catalysts for societal discourse. Research has shown that media content with high production quality and strong narrative elements fosters emotional engagement, which can enhance message retention and stimulate critical thinking among viewers \citep{borkiewicz2019cinematic,arroio2010,franconeri2021science,lee2022affective}. This underscores the importance of persuasive storytelling and high-quality visuals, as they play a critical role in bridging the gap between scientific content and public understanding.
Moreover, the recurring themes of engagement and learning reflect the capacity of documentaries to create shared cultural moments that go beyond individual experiences. Studies suggest that collaborative viewing and subsequent discussions can amplify the influence of documentaries, encouraging viewers to reflect on social norms and consider alternative perspectives \cite{whiteman2009documentary,hirsch2007documentaries}.

Finally, the themes show the potential for documentaries to reinforce or challenge preconceived notions about science. For instance, when technical visualization aspects are well-executed, they not only enhance comprehension but also demystify complex scientific processes, making science appear more accessible and inclusive \cite{yang2020power}. These findings align with the broader literature emphasizing that media's impact is mediated not just by content but also by how that content resonates with and activates viewers' values, beliefs, and aspirations \cite{rezapour2017classification}.




\section{Conclusion and Future Work}
This paper analyzes the impact of scientific films using user-generated reviews and measures the impact using a novel taxonomy of change and engagement. 
Beyond the immediate goals of \avlshort\ the results of our best model as outlined in \autoref{tab:Classification Result Sentiment} and \autoref{section:classification} indicate our model's efficiency in predicting impact and sentiment which can be useful in future productions assessment by the \avlshort\ and other teams to assess the impact of their produced works. Additionally, the results of our generalizability analysis presented in \autoref{tab:Hubble_Prediction} and \autoref{section:generalizability} indicate the model should be applicable to a wide range of scientific CSV-style documentaries.
Overall, our results underscore the multifaceted impact of scientific documentaries. They engage audiences emotionally, provoke intellectual curiosity, and influence perceptions, making them a powerful medium for science communication. Our findings contribute to a growing body of evidence on the effectiveness of media in bridging the gap between science and the public, emphasizing the value of leveraging documentaries to achieve broader educational and societal goals. Future research could explore the long-term impacts of such engagement, particularly in terms of sustained interest in science and changes in behavior or policy advocacy.

The \avlshort\ expressed interest in tying specific learning goals to sentiment and impact categories and/or impact themes.
As learning goals are often developed to guide the creation of these scientific documentaries and are integral to grant funding, our large-scale quantitative analysis of the impact of these documentaries would be a useful complement to the smaller-scale audience-based feedback the \avlshort\ currently collects.
While our model was trained and tested on Amazon comments, our collaboration also has access to YouTube comments for a subset of the analyzed films. Future work will include an analysis of these comments, however, there is evidence of a bias toward more negative reviews on YouTube \citep[e.g.][]{ardestani2024youtube}, making such a comparison a test of the inter-platform generalizability of our model.


\section{Limitations and Ethics Statement}
Our data collection is limited only to English reviews, impacting the generalizability of our findings.
We also limited the scope of our study to only one platform, i.e., Amazon. This can enable platform-specific biases to influence the findings, and classification performance and cause potential inaccuracies.
In addition, media impact reviews are self-reported and collected only once. Without access to follow-up reviews or more nuanced, contextualized reports, we cannot be certain of the persistence and durability of such reported impacts. 
Finally, we acknowledge that LLM-based classifiers often make mistakes in predicting the best label, resulting in misclassified reviews. Future improvements in these models and better prompt engineering can enhance the performance of LLM classifiers.

\begin{acks}

The authors thank the work of Alistair Nunn, Rishabh Sharma, and Shashwat Mann for their annotation and analysis consultations. The authors additionally thank the members of the Advanced Visualization Laboratory at the National Center for Supercomputing Applications for their collaboration on this work.  Finally, the authors thank the referees of this work for their valuable input.
\end{acks}

\bibliographystyle{ACM-Reference-Format}
\bibliography{aaai22}


\appendix
\section{Annotation Platform}
\begin{figure*}[t]
\centering
\includegraphics[width=0.6\textwidth]{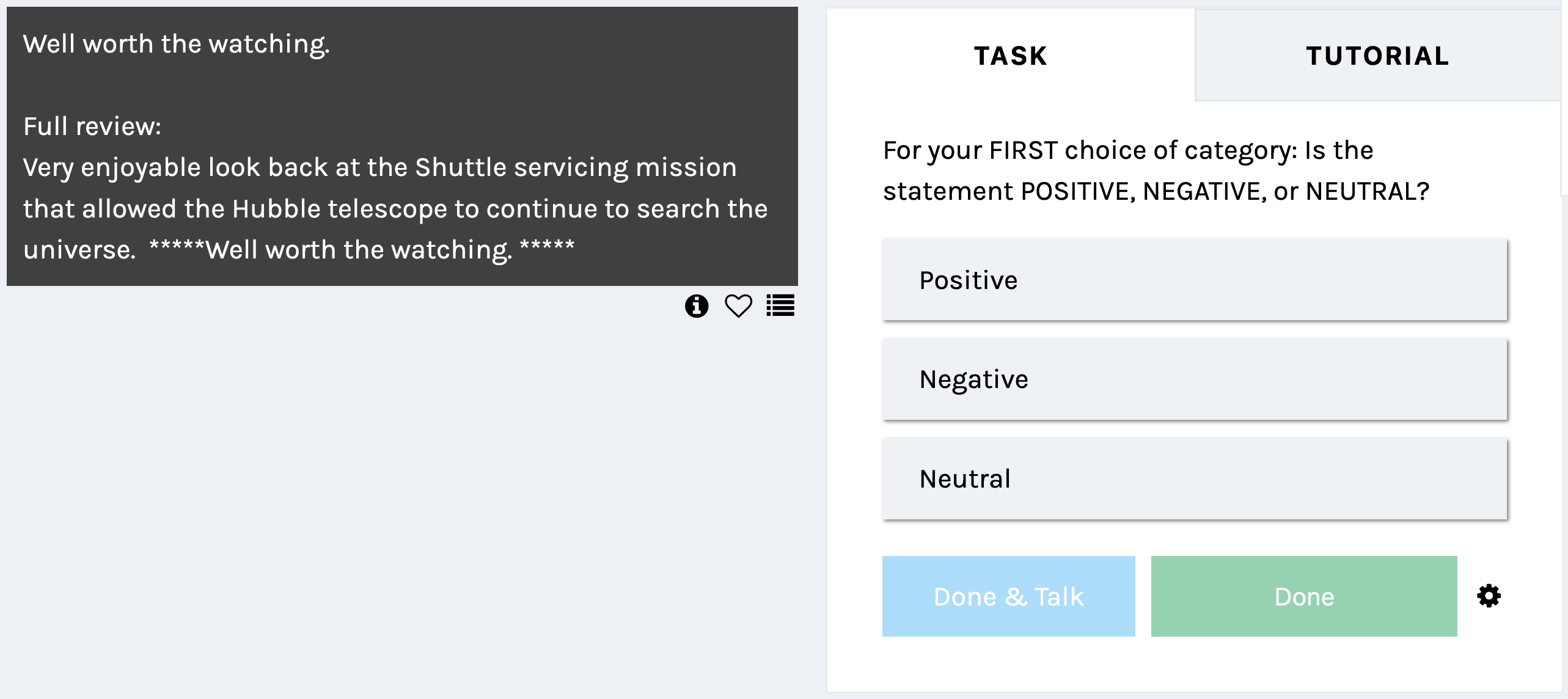} 
\caption{First stage of annotation process in which users are instructed to select their first choice for the sentiment category for each sentence. Each sentence (top) is shown in the context of the full review (bottom). The annotator moves to the next stage of annotation with the green ``Done'' button (or blue ``Done \& Talk'' button to post comments in the project forum).  More information (comment URL and film title) is accessed with the circled ``i'' Info button and annotation can be saved in the user's personal collection for later reference with the heart Favorite button.}
\label{fig:first_step_annotation}
\end{figure*}

We used the Zooniverse for the annotation process. 
Zooniverse boasts over one million active members collaborating on hundreds of projects ranging from the space sciences to transcription of historical documents\footnote{\url{https://www.zooniverse.org/about}}. 
Here, ``first choice'' is defined as the strongest sentiment expressed in the statement. If there were two equally strong sentiments expressed, then the sentiment that appeared first was selected as the ``first choice'' (the analysis of ``second-choice'' selections is relegated to a future paper).
Once the user made their selection, they could select ``Done'' to move to the next stage of the annotation (green button in \autoref{fig:first_step_annotation}) or ``Done \& Talk'' to ask questions in the main forum (blue button in \autoref{fig:first_step_annotation}).
The annotator was then prompted to repeat the process with their first choice for impact category (categories listed in the first column of \autoref{tab:impact_categories}).

\section{Prompt design}\label{appendix}
\begin{figure*}[ht]
    \centering
    \includegraphics[width=0.7\textwidth] {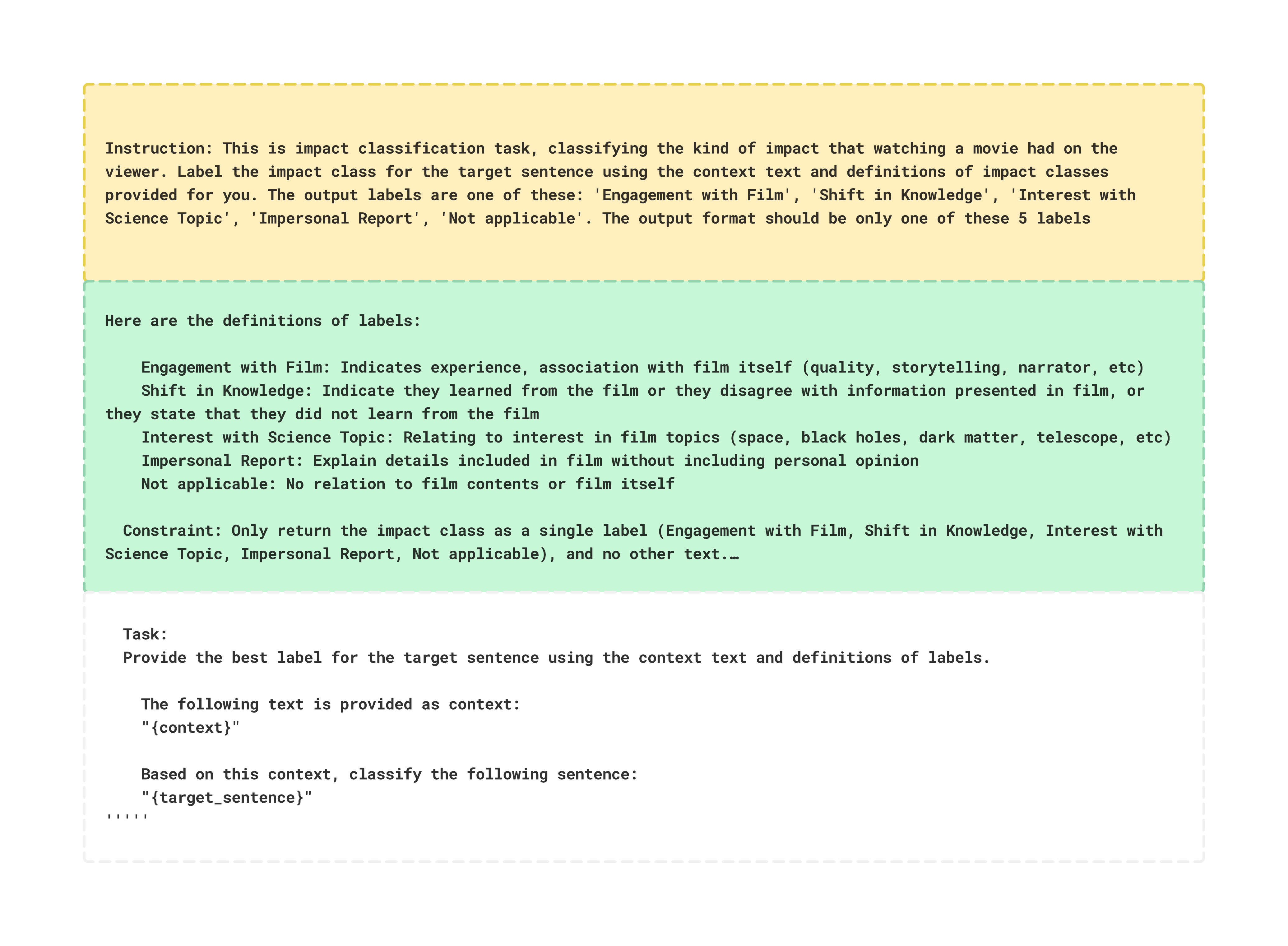}
    \caption{Prompt used for prediction of impact categories}
    \label{Prompt}
\end{figure*}
Figure \ref{Prompt} illustrates the prompt we used to develop LLM classifiers for impact categories that are instructed to label a target sentence, using full reviews as context. The prompt begins with an instruction of the classification task, followed by the provided definitions of impact labels, and feeding both the context and target sentence to the prompt at the end.

\end{document}